\documentclass{aa} 
\usepackage{graphicx,natbib} 
\usepackage{txfonts} 
\bibpunct{(}{)}{;}{a}{}{,} 
 
\def\water{H$_2$O~}
\def\kms{km~s$^{-1}$}

\newcommand{\nc}{\newcommand}

\nc{\RAJ}[4]{$\alpha(J2000) = {#1}^{\rm h}{#2}^{\rm m}{#3}\fs{#4}$}
\nc{\DecJ}[4]{$\delta(J2000) = {#1}\degr {#2}\arcmin {#3}\farcs{#4}$}

\begin{document} 
 
\title{From the ashes: JVLA observations of water fountain nebula candidates show the rebirth of IRAS 18455+0448}
 
\author{Vlemmings, W. H. T.\inst{1} 
\and Amiri, N.\inst{2}
\and van~Langevelde, H.J.\inst{3,4}
\and Tafoya, D.\inst{5}}

\institute{Department of Earth and Space Sciences, Chalmers University of Technology, Onsala Space Observatory, SE-439~92 Onsala, Sweden
\email{wouter.vlemmings@chalmers.se} 
\and Center for Astrophysics and Space Astronomy, Department of Astrophysical and Planetary Sciences, University of Colorado, 389 UCB, Boulder, CO, 80309-0389, USA
\and Joint Institute for VLBI in Europe, Postbus 2, 7990 AA,
Dwingeloo, The Netherlands
\and Sterrewacht Leiden, Leiden University,
Postbus 9513, 2300 RA, Leiden, The Netherlands
\and Centro de Radioastronom\'\i a y Astrof\'\i sica, UNAM, Apdo. Postal 3-72 (Xangari), 58089 Morelia, Michoac\'an, M\'exico
}
\date{received , accepted} 
\authorrunning{Vlemmings et al.} 
\titlerunning{The reborn water fountain IRAS 18455+0448} 
 
\abstract {The class of water fountain nebulae is thought to represent the stage of the earliest onset of collimated bipolar outflows 
during the post-Asymptotic Giant Branch phase. They thus play a crucial role in the study of the formation of bipolar Planetary
  Nebulae (PNe). To date, 14 water fountain nebulae have been identified. The identification of more sources in this unique stage
  of stellar evolution will enable us to study the origin of bipolar PNe morphologies in more detail.}  
  {Water fountain candidates can be identified based on the often double peaked 22~GHz \water maser spectrum with a large separation 
  between the maser peaks (often $>100$~\kms). However, even a fast bipolar outflow will only have a moderate velocity extent in its maser 
  spectrum when located close to the plane of the sky. In this project we aim to enhance the water fountain sample by identifying objects 
  whose jets are aligned close to the plane of the sky.}  
  {We present the results of seven sources observed with the Jansky
    Very Large Array (JVLA) that were 
    identified as water fountain candidates in an Effelsberg 100-m
    telescope survey of 74 AGB and early post-AGB stars. }  
  {We find that our sample of water fountain candidates displays strong variability in their 22~GHz \water maser spectra. The JVLA 
  observations show an extended bipolar \water maser outflow for one
  source, the OH/IR star IRAS~18455+0448. This source was 
  previously classified as a dying OH/IR star based on the exponential decrease of its 1612~MHz OH maser and the lack of \water masers. 
  We therefore also re-observed the 1612, 1665, and 1667~MHz OH masers. We confirm that the 1612~MHz masers have not reappeared 
  and find that the1665/1667~MHz masers have decreased in strength by several orders of magnitude during the last decade.
  The JVLA observations also reveal a striking asymmetry in the red-shifted maser emission of IRAS~19422+3506.}  
  {The OH/IR star IRAS~18455+0448 is confirmed to be a new addition to the class of water fountain nebulae. Its kinematic age is $\sim70$~yr, 
  but could be lower, depending on the distance and
  inclination. Previous observations indicate, with significant
  uncertainty, that IRAS~18455+0448 has a surprisingly low mass
  compared to available estimates for other water fountain nebulae. The available historical OH~maser
  observations make IRAS~18455+0448 unique for the study of water fountain nebulae and the launch of post-AGB bipolar outflows. The
  other candidate sources appear high mass-loss OH/IR stars with partly radially beamed \water masers.}
 
\keywords{Stars: (post-)AGB, stars: evolution, outflows, masers} 
 
\maketitle 
 
\section{Introduction} 

The origin of the bipolar morphologies observed in a large number of
young planetary nebulae (PNe) and post-asymptotic giant branch
(AGB)/pre-PNe sources is a long-standing question. Various mechanisms
explaining the observed morphologies have been invoked
\citep{Balick02}. Of these, collimated outflows launched during the late
AGB or early post-AGB phase has been suggested as the most direct
explanation for the formation of bipolar PNe
\citep[e.g.,][]{Sahai98}. The origin of these collimated outflows, or
bipolar jets, is however still unclear. The collimation of an outflow
due to a magnetic field, similar to what is observed for
proto-stars, is a promising candidate, even though such a field might
require a binary companion or massive planet to be maintained
sufficiently long \citep[e.g.,][]{Garcia14}. Alternative
common-envelope evolutionary scenarios in which a binary companion
becomes engulfed by the expanding AGB star could also provide an
explanation \citep[e.g.,][]{Nordhaus07}.

In order to constrain the launching mechanisms  of bipolar
  outflows in late-type stars, it is important to find the youngest
bipolar sources where the ejection of material has begun just
  recently. These are often still in, or just after, the AGB phase of
their evolution. For these sources it can be possible to study
  simultaneously both the bipolar outflows as well as the remnants of
their AGB envelopes. The youngest of the bipolar pre-PNe/post-AGB
stars are the class of water fountain sources. These display \water
maser emission at velocities that can reach up to several hundred
\kms. This is significantly beyond that of the regularly expanding
envelopes that can sometimes still be traced in OH masers, with
typical expansion velocities of $\sim15$~\kms~ \citep{Likkel92}.
Interferometric observations of these sources have revealed that they
posses tightly collimated jets that excite \water masers at the
shocked interface ahead or around them \citep{Imai07}. These jets have
kinematic ages of only a few tens to hundreds of years and synchrotron
and maser polarization observations indicate that at least a few of
them are collimated by a strong magnetic field
\citep{Vlemmings06,PerezSanchez13}. The Galactic scale height of the
  water fountain population, the heavy circumstellar extinction and observations of
  $^{12}$CO and $^{13}$CO appears to indicate that their progenitor
  stars are fairly massive \citep[$>4$~M$_\odot$][]{He08, Suarez08,
    Imai12, Rizzo13}.  None of the water fountains has had a binary
companion directly confirmed, although jet-precession in some of them
could be due to an embedded companion \citep{Imai07, Yung11}.

Only 14 water fountains are known to date. Since their detection is
often related to the observations of \water masers spread over a very
large velocity range, detection is necessarily biased towards bipolar
sources closest to the line of sight. In order to identify water
fountain nebulae whose jets lie close to the plane of the sky, a
combination of single dish monitoring and interferometric observations
is required. Here we present JVLA and Effelsberg observations of a
number of candidate water fountain sources with a smaller spread of
their \water masers. The source sample, observations and data
reduction are presented in \S~\ref{obs}.  The resulting maser spectra
and \water maser variability are described in \S~\ref{results}, and in
particular the results of the confirmed water fountain
IRAS~18455+0448 are discussed in \S~\ref{discussion}.

\begin{figure*}
\centering
\resizebox{0.9\hsize}{!}{\includegraphics{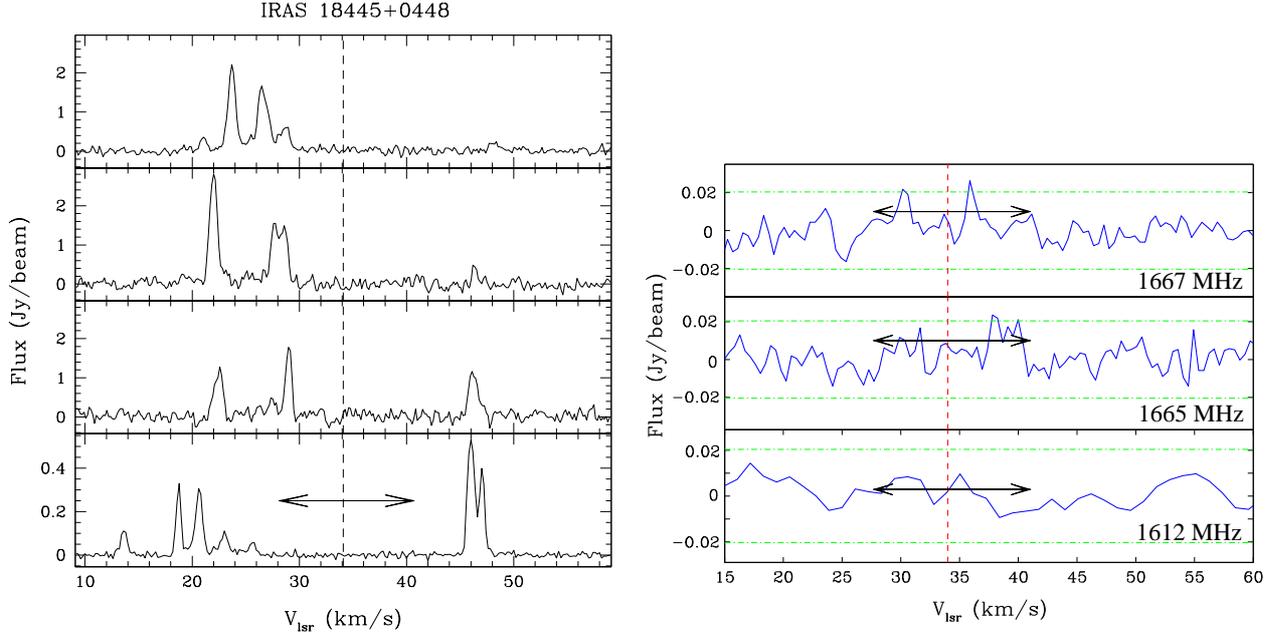}}
\caption{(left) Spectra of the \water masers of IRAS~18455+0448
 observed with the Effelsberg 100-m telescope (top three panels) and
 with the JVLA (bottom panel) at four different epochs (with
 different flux scales). From top to
 bottom, the epochs are 12-13 November 2009, 22-23 March 2011, 14
 April 2011 and 6 June 2013. The arrows in the bottom panel indicate
 the spread of the previously observed OH masers and the vertical
 dashed line indicates the systemic velocity based on SiO maser
 observations. (right) Confirmation of the disappearance of the OH
 masers at 1612, 1665 and 1667~MHz. The masers were observed with
 Effelsberg at 22 February 2010 (1665 and 1667~MHz) and 27 March 2010
 (1612~MHz). The vertical dashed line indicated the systemic velocity
 and the horizontal dotted-dashed lines the $3\sigma$ noise
 limits. The arrows indicate the spread of OH masers observed by
 \cite{Lewis01}. } \label{irasspec}
\end{figure*}

\begin{figure*}
\centering
\resizebox{0.85\hsize}{!}{\includegraphics{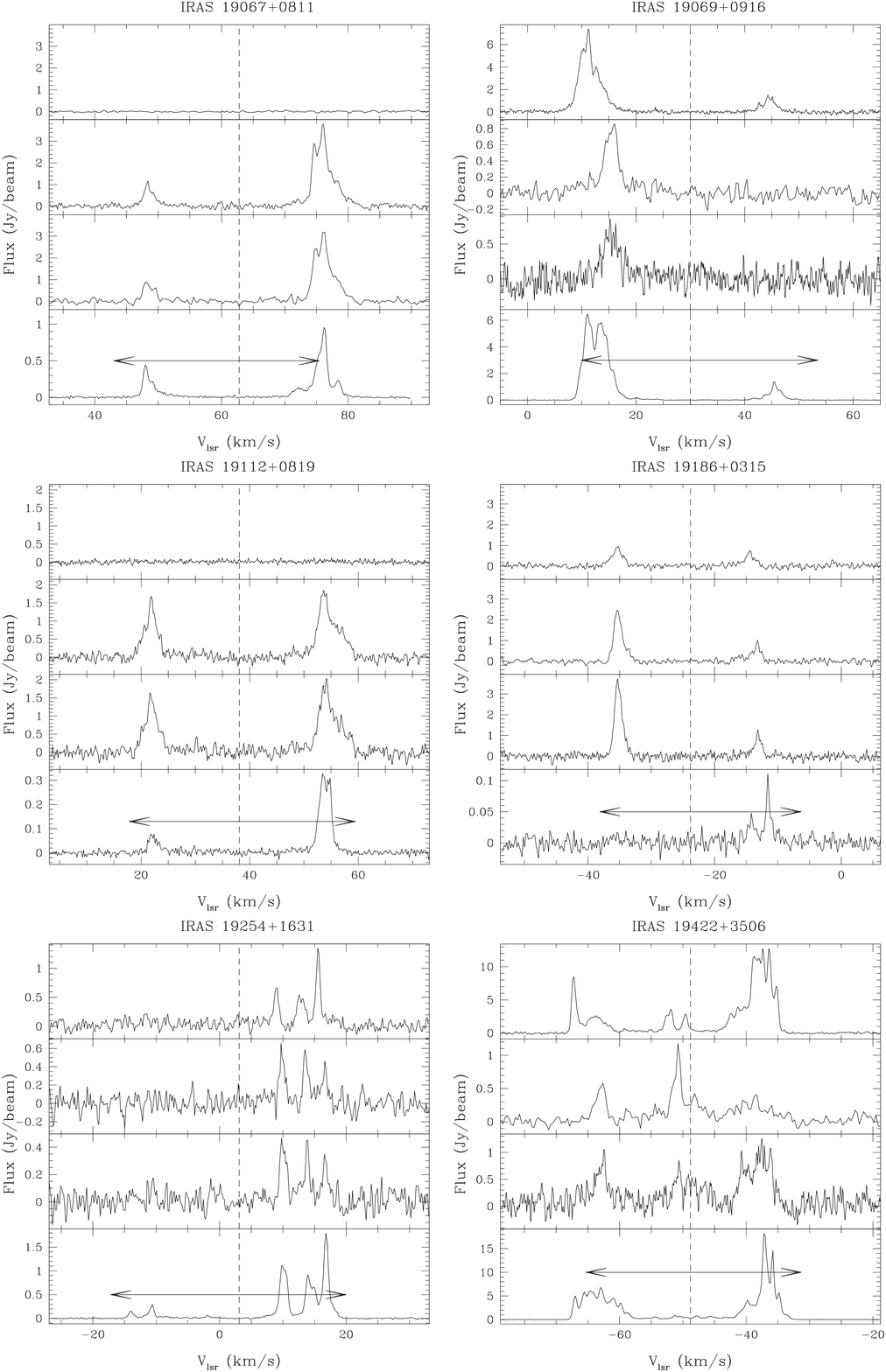}}
\caption{Same as Fig.~\ref{irasspec}(left) for the other six sources in our
sample. We note the difference in flux scale for several of the observational
epochs. The vertical dashed line indicates, where available, the literature
averaged SiO~maser velocity, otherwise the systemic velocity derived
from the OH masers is used (see Table.~\ref{sources}).} \label{specs}
\end{figure*}

\section{Observations} 
\label{obs}

\subsection{The sample}

The sources in our sample were originally observed as part of the
observational campaign presented and discussed in \cite{AmiriThesis}. In that work,
a sample of 74 late AGB or early post-AGB objects was observed in an
attempt to detect their \water maser emission. 
In this
work, we focused on seven sources that were either newly discovered or
have distinct double or multipeaked \water maser emission with a
velocity spread close to that of the OH masers. The sources are listed
in Table~\ref{sources}. One of the sources identified in
  \cite{AmiriThesis} as a water fountain candidate,IRAS~18455+0448,
  was also proposed as a water fountain based on later Effelsberg
  observations by \cite{Yung13}. As earlier observations 
  indicated an exponential decline in the 1612~MHz OH maser line for
  this source \citep{Lewis01}, we also performed new 1612, 1665, and
  1667~MHz OH maser observations for IRAS~18455+0448.

\subsection{Effelsberg observations}
Observations of the 22.23508 GHz \water masers in our sample were done
with the Effelsberg telescope over three epochs: 12 and 13 November
2009, 22 and 23 March 2011, and 14 April 2011. For the first two epochs
we used the 1.3 cm prime focus receiver in spectral line mode. The
1.3 cm focus secondary VLBI receiver was used for the third epoch
observations.

The full width half maximum (FWHM) of the Effelsberg telescope is 40\farcs2 
at the maser frequency. Using the FFT spectrometer with 16384
channels and a bandwidth of 100 MHz, equivalent to a velocity coverage of 1363 km~s$^{-1}$, 
centered at the stellar velocity, the resulting spectral resolution is 0.08
km~s$^{-1}$. The observations were made in position switching mode
with a cycle of 1 min, sufficient to compensate for atmospheric
fluctuations. The observing time for each source was $8-16$~min at epoch
1, $8$ min at epoch 2 and $12-24$~min at epoch 3. The rms noise for the
first, second and third epochs corresponds to $0.02-0.18$, $0.06$ and
$0.07-0.3$ Jy, respectively.

For the OH maser observations of IRAS~18455+0448, we used the 21/18 cm
primary focus receiver. At the OH maser frequency the FWHM of the
telescope corresponds to $\sim$470 arcsec. Mainline OH maser
observations at 1665.4018 and 1667.359 were performed on 22 Feb 2010
with a bandwidth of 20 MHz and 16384 channels which provides a channel
spacing of $\sim$0.2 km~s$^{-1}$. The total on source observing time
was 2 hr and 26 minutes. The observations of 1612 MHz masers were made
on 27 March 2010 with a bandwidth of 100 MHz and 16384 channels
corresponding to a channel spacing of $\sim$1 km~s$^{-1}$. The total
on source observing time was 72 minutes. The rms noise in channels
free of emission is $\sim$7 mJy for both epochs.

The data reduction was performed using the Continuum and Line Analysis
Single-dish Software (CLASS) that is part of the GILDAS
package\footnote{http://www.iram.fr/IRAMFR/GILDAS}.  We obtained the
raw spectra in units of temperature counts. The spectra were converted
to units of antenna temperature (T$_{A}$) by applying the calibration
noise temperature. We also corrected the spectra for atmospheric
opacity and the gain-elevation effect, the measurements of which
  were provided by the observatory. Finally, the spectra were
converted to Jy units by dividing the spectra by the sensitivity of
the telescope at the maser frequency.

\subsection{JVLA observations}

In order to improve the positional accuracy of our sources for later
VLBI observations and investigate the location and extent of the
\water masers, our sources were observed with the JVLA in
  C-configuration on 6 June 2013 (program 13A-088). We used a total
bandwidth of 32~MHz centered at the \water maser rest frequency in
dual polarization. The original 8192 channels were averaged to produce
a data set with 4096 channels of $\sim0.1$~\kms~ width. The seven
sources were observed interleaved with four phase calibrators
(J2007+4029, J1851+0035, J1922+1530, and J1925+2106) during a total
observing time of $3~$hr. Bandpass calibration was done using
J2007+4029 and we performed periodic pointing observations at X-band.

The data calibration was performed using the Common Astronomy Software
Application (CASA). As no suitable primary flux calibrator was visible
during our short observing run, we rely on archive fluxes of our phase
calibrators to determine the absolute flux of our sources. Based on
this we conclude that our fluxes are uncertain at the level of $25\%$.
Although the positions of the masers in our sample were determined by
directly fitting the uv-data, we also produced images and spectra
further averaging to $0.2$~\kms~ channels. The typical rms in our
spectra is $7.2$~mJy/beam and the typical beamwidth is
$\sim1.2"\times1.1"$. A direct comparison between the fluxes
  obtained from single-dish Effelsberg and from interferometric JVLA
  observations is possible, since \water masers with typical sizes of only a
  few milliarcseconds are too
  compact to be resolved out using the JVLA in C-configuration.

\begin{table*}
\caption{Observed sources}
\begin{tabular}{|l|cc|c|c|c|c|c|c|}
      \hline
    Source        & $\alpha_{\rm J2000}$ & $\delta_{\rm J2000}$ & $D^a$ &
     $V_{\rm lsr}^b$ & $\Delta V_{\rm OH}$ & $\Delta V_{\rm H_2O}^c$ &
     $\Delta_{\rm red-blue}$ & refs.\\
             & $hh~mm~ss$ & $^\circ~'~"$ & kpc & \kms & \kms & \kms &
             mas &\\
   \hline
   \hline
IRAS~18455+0448 & 18~48~02.30 & +04~51~30.446 & 2.1/11.2 & 34.1 & 12.5 & 38.7 &
$49\pm15$ & 1\\
IRAS~19067+0811 & 19~09~08.31 & +08~16~33.802 & 3.7/8.6 & 62.8/59.2 &
32.5 & 31.8 & $13\pm12$ & 2,3\\
IRAS~19069+0916 & 19~09~19.25 & +09~21~11.529 & 2.0/10.1 & 30.0/31.7 &
43.6 & 38.1 & $8\pm6$ & 4,5\\
IRAS~19112+0819 & 19~13~37.32 & +08~24~52.489 & 2.4/9.8 & 38.6 & 41.8
& 33.6 & $18\pm45$ & 6\\
IRAS~19186+0315 & 19~21~11.71 & +03~20~57.800 & 14.7 & -23.8/-22.0 &
31.8 & 28.0 & - & 7,5\\
IRAS~19254+1631 & 19~27~42.03 & +16~37~24.182 & 10.3 & 3.1/1.9 & 37.2
& 29.0 & $19\pm15$ & 3,8\\
IRAS~19422+3506 & 19~44~07.00 & +35~14~08.207 & 10.3 & -48.8/-48.4 &
34.0 & 34.7 & $4\pm5$ & 7,9\\ 
   \hline
\multicolumn{9}{l}{$^a$ near/far kinematic distance \cite[using][]{Reid09}.}\\
\multicolumn{9}{l}{$^b$ literature averaged SiO and OH systemic velocities. If both are the
same or SiO is unavailable only a single value is given.}\\  
\multicolumn{9}{l}{$^c$ based on the JVLA observations, except for
IRAS~19186+0315, for which the third Effelsberg epoch was used.}\\
\multicolumn{9}{l}{References: 1-\citet{Lewis01},2-\citet{Kim10},3-\citet{Engels86},4-\citet{Nakashima03},5-\citet{Eder88}}\\
\multicolumn{9}{l}{6-\citet{Chengalur93} ,7-\citet{Kim13},8-\citet{tLH89},9-\citet{Lewis97}}
\end{tabular}
\label{sources}
\end{table*}

\section{Results}
\label{results}

\subsection{\water maser spectra}

The Effelsberg and JVLA \water maser spectra for IRAS~18455+0448 and
those for the other six sources are shown in Fig.\ref{irasspec} (left)
and Fig.~\ref{specs} respectively. All seven sources display
significant variability in their \water maser emission between the
four epochs. In particular, we find strong variability for IRAS
19067+0811 and IRAS 19112+0819, which were undetected in our first
Effelsberg epoch, while for IRAS 19254+1631 only the JVLA
  observations reveal clear blue-shifted emission. Moreover, IRAS 19069+0916 and IRAS 19422+3506,
display variability of almost two orders of magnitude between
different epochs. The strong variability could also be
the cause of the initial non-detection of IRAS~18455+0448 (see however
\S~\ref{irasdisc}), IRAS~19112+0819 and IRAS~19254+1631 in earlier
observations by \citet{Engels96}.

In all cases except IRAS~18455+0448, the \water maser velocity spread
is similar to that of the OH masers. While all of the sources
  display at least a prominent double peaked spectrum,
  IRAS~18455+0448, IRAS~19254+1631 and IRAS~19422+3506 also display
  several other strong maser features across their spectrum.

\begin{figure}
\centering
\resizebox{0.9\hsize}{!}{\includegraphics{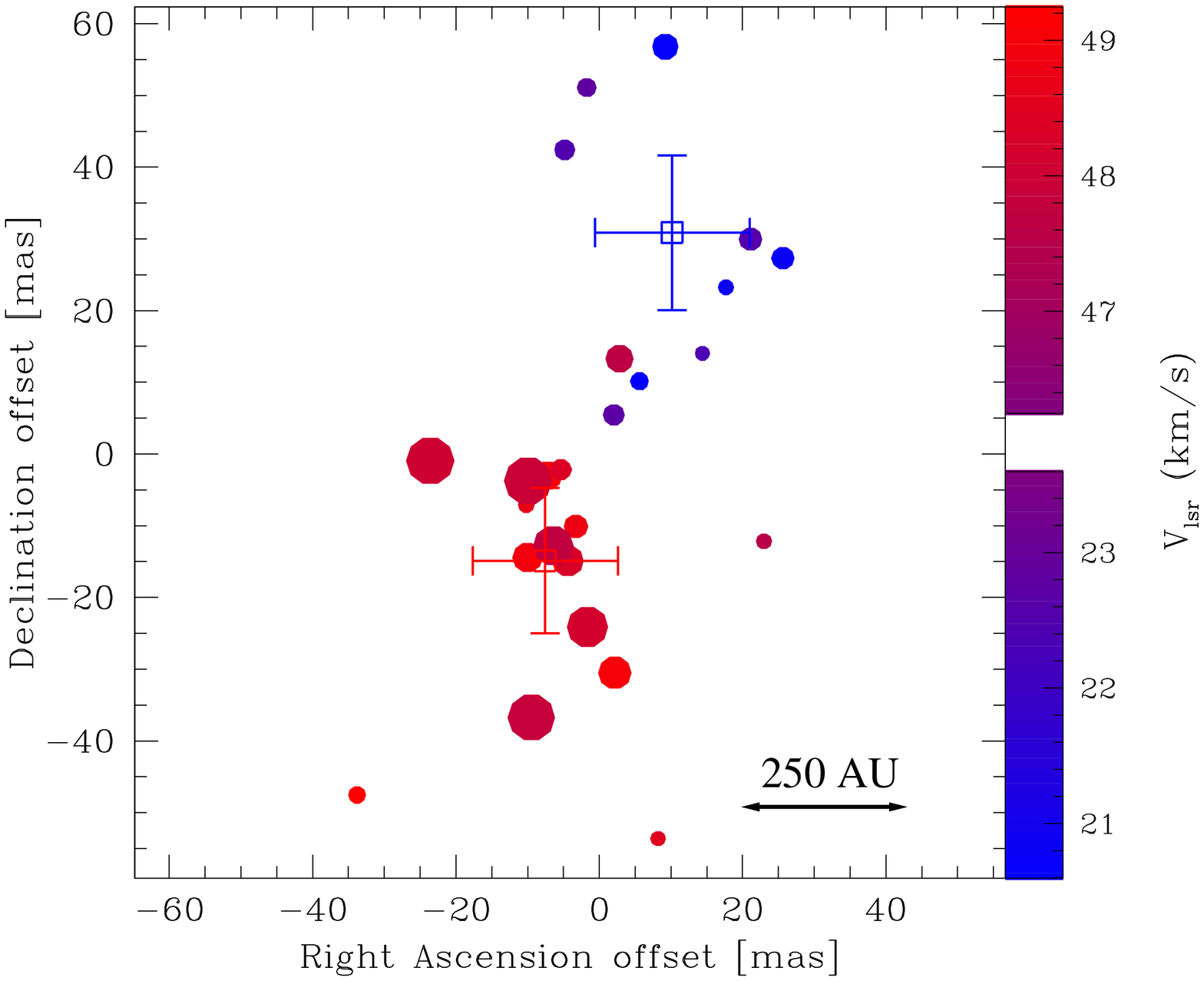}}
\caption{(The \water maser distribution around IRAS~18455+0448
  observed with the JVLA. Fits to the centroid position of individual
  maser velocity channels are potted scaled based on their
  flux and color coded according to velocity. Only masers detected at
  a signal-to-noise-ratio $>15$ are plotted. The positional
  uncertainties on the individual symbols range from $\sim15$~mas for
  the brightest to $\sim30$~mas for the weakest features. The error weighted mean positions of the blue- and
  red-shifted masers with corresponding error bars are also indicated. The scale,
  assuming a distance of $11.2$~kpc is indicated in the bottom right
  corner. } \label{iras18455}
\end{figure}

\subsection{OH masers of IRAS~18455+0448}

The OH maser spectra of IRAS~18455+0448 are shown in
Fig.~\ref{irasspec} (right). The OH 1612~MHz maser is not detected
after being measured to have a peak intensity of $\sim40$~mJy in 2000
\citep{Lewis01}. In that epoch, the 1665~ and 1667~MHz OH masers were
observed to have peak fluxes of $\sim450$~mJy and $\sim730$~mJy
respectively. In our current observations we only tentatively detect
possible peaks of the 1665~ and 1667~MHz OH masers at $25\pm7$~mJy
(at $V_{\rm lsr}\approx38$~\kms) and $22\pm7$~mJy (at $V_{\rm
    lsr}\approx30$~and~$36$~\kms) 
respectively. The relative velocity, with respect to the star, of
these potential maser peaks are less than what was measured in 2000
 as indicated in Fig.~\ref{irasspec}(right) by the arrow. The main-line masers have thus also faded by more than
an order of magnitude over the last decade.

\subsection{JVLA observations of the \water maser components}

\begin{figure*}
\centering
\resizebox{0.9\hsize}{!}{\includegraphics{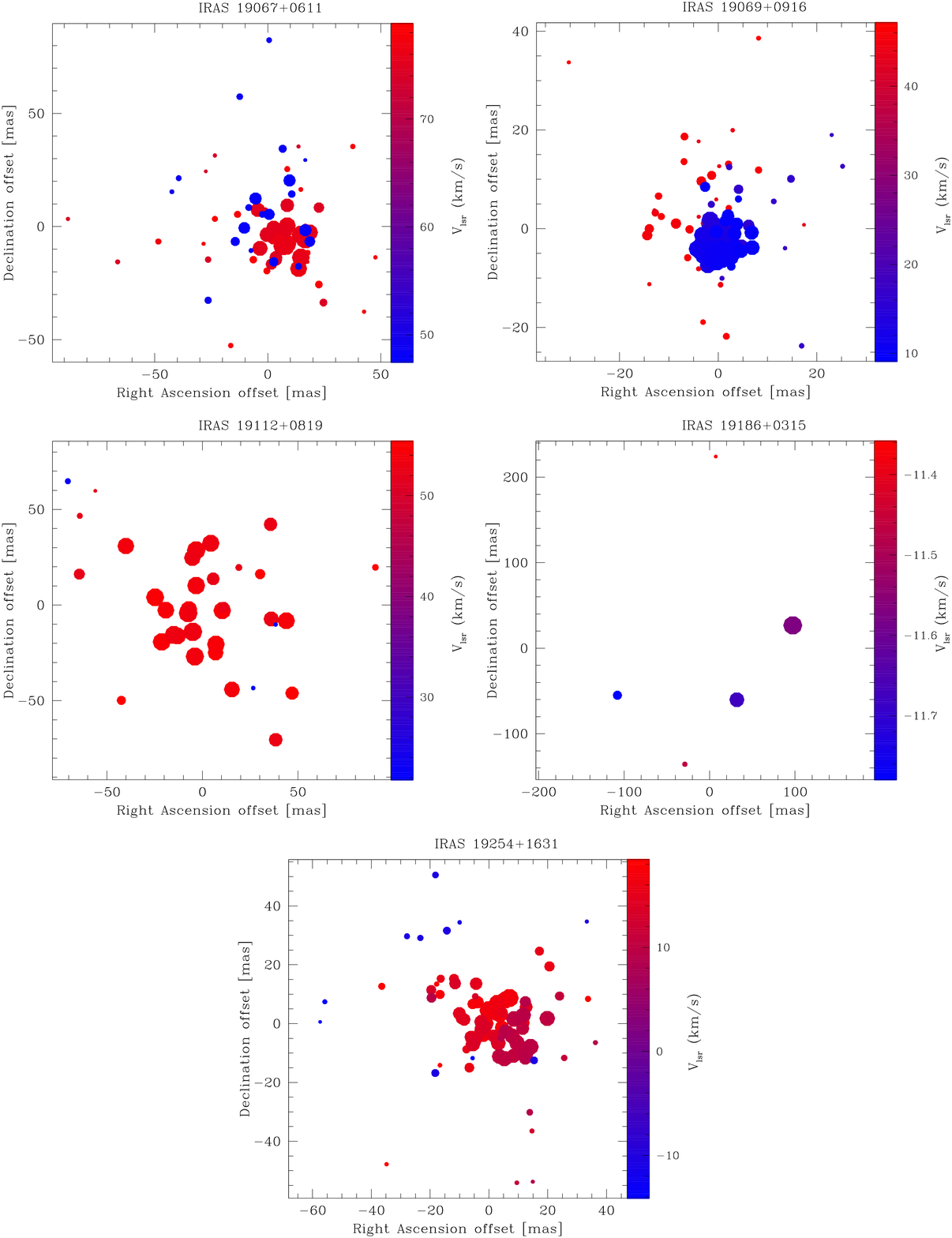}}
\caption{(left) Same as Fig.~\ref{iras18455} for the remaining
  source in the sample. Depending on maser flux, different SNR cutoffs
  are used. For IRAS~19067+0611 all channels with SNR $>5$ are
  plotted and error bars range from $\sim30$~to~$90$~mas. For
  IRAS~19096+0916+0819 we use SNR $>25$ and error bars range from
  $\sim5$~to~$18$~mas. For IRAS~19112+0819, SNR $>5$ and error bars
  range from $\sim25$~to~$90$~mas. For IRAS~19186+0325, SNR $>5$ and
  error bars range from $\sim75$~to~$90$~mas. Finally, for
  IRAS~19254+1631,  SNR $>10$ and error bars range from $\sim10$~to~$45$~mas.} \label{all}
\end{figure*}

Our observations with the JVLA allowed us to
determine more accurate positions and to investigate the
offset between the red- and blue-shifted masers. Although the
$\sim1"$ JVLA resolution is not sufficient to directly resolve the
individual components, the strength of the masers allows us to reach
positional accuracies of a few to a few tens of milliarcseconds for
the maser positions (considering $\theta\approx{\rm Beam}/(2\times{\rm
  SNR})$), where SNR is the signal-to-noise-ratio. We determined the maser positions by performing fits in the
(u,v)-plane for each velocity channel assuming a single, unresolved,
component \citep[using the CASA code UVMULTIFIT][]{MV14}. Although in
the spectra several individual features can be identified, these
likely still consist of individual maser components blended in
velocity. However, fitting multiple maser components within the 
relatively large JVLA beam would lead to poorly defined fits. The position errors
determined from the u,v-fitting of a single component per velocity
channel will, besides the aforementioned dependence on SNR and
beam-size, contain the uncertainty related to the potential existence
of multiple components in the beam. Component maps for IRAS~18455+0448
and IRAS~19422+3506 are shown in Figs.~\ref{iras18455}
and~\ref{iras19422} respectively. The remaining sources are shown
in Fig.~\ref{all}. For all sources (except
IRAS~19186+0315 where only red-shifted emission was detected), we
determined the offset between the red- and blue-shifted emission by
determining a SNR weighted average of the channel component
positions. The results of this analysis are presented in
Table,~\ref{sources}. That same table lists the JVLA observed spread
of the \water masers velocities ($\Delta V_{\rm H_2O}$) and the
centroid positions determined by SNR weighted averaging of all maser
channels. As can be seen in Fig.~\ref{iras18455}, IRAS~18455+0448 is
the only source for which a significant off-set, of $49\pm15$~mas, is
measured.

\section{Discussion} 
\label{discussion}

\subsection{Nature of IRAS~18455+0448}
\label{irasdisc}

IRAS~18455+0448 is the only source of the sample for which
a clear separation is seen between the red- and blue-shifted masers
(Fig.~\ref{iras18455}) and for which the \water maser velocity spread
is more than twice that of the OH masers. It was also one of the three
sources in our Effelsberg survey that displayed \water masers that
were not detected in the previous survey by \citet{Engels96} who 
reported an rms level of $0.2$~Jy in 1990.

Based on the exponential fading of the 1612~MHz OH masers between 1988
and 2000, \citet{Lewis01} suggested the source as the prototype of a
'dying' OH/IR star. In their preferred model, the mass loss has
gradually declined to close to 0. Considering the dust velocity ($V_{\rm e,
  dust}$) is larger than that of the gas ($V_{\rm e, gas}$), the
expansion of both will differ. This results in a gradual decrease in the dust
column density that shields the OH from the interstellar UV. The OH
will thus be photo-dissociated more rapidly, thereby decreasing the
column density in the maser region. This effect will become apparent
after a time that is proportional to the expansion timescale from the
photosphere to the OH maser region and the ratio $V_{\rm e,
  gas}/V_{\rm e, dust}$. This is on the order of $1000$~yr for
IRAS~18455+0448. As the satellite OH line has the strongest dependence
on column density, it will initially be affected before the main line
masers. In the framework of the above model, an abrupt termination of
the mass loss would not produce the exponential decline in maser flux,
nor would it affect the 1612~MHz masers before the main line
masers. Thus, the model described by \citet{Lewis01} is supported by
the fact that now, ten years later, the main line masers have also
decreased by at least an order of magnitude (Fig.~\ref{irasspec},
right).

However, the model described above does not take the effect of maser
pumping into account. As discussed in \citet{Gray05}, the response of
the 1612~MHz OH maser flux could also be much more rapid if one takes
into account a loss of radiative pumping efficiency after dust
production has halted. In that case, a more rapid decline of mass loss
can, already within a few tens of years, result in the observed
exponential decay of the maser flux. The significant time-lag of at
least a decade between the decay of the 1612~MHz and the 1665/67~MHz
masers is however not as easily explained, as in this model the main
line masers would be the first to disappear.

In the model of significantly decreasing mass loss, the \water masers
cannot be excited in the inner circumstellar envelope. This suggests
that the fast \water masers are excited in the interaction between a
fast outflow and the remnant envelope as is found for water fountain
nebulae. Our JVLA maser component map in Fig.~\ref{iras18455} confirms
that the red- and blue-shifted masers are separated by $\sim50$~mas
($\sim560$~AU at $11.2$~kpc). 
Considering the separation, both
spatially and spectrally, and elongated morphology of the red- and
blue-shifted masers, we can conclude that IRAS~18455+0448 belongs to
the class of water fountain sources.

We can use the measured separation to make a crude estimate the kinematic age of
IRAS~18455+0448. Assuming a constant fast outflow velocity, the age is
given by
\begin{equation}
\left[\frac{t_{\rm kin}}{\rm yr}\right] = 4.74\left[\frac{D}{\rm kpc}\right]\left[\frac{\Delta_{\rm red-blue}}{\rm mas}\right]\left[\frac{\Delta V_{\rm
      H_2O}}{\rm km~s^{-1}}\right]^{-1}\tan\left[\frac{i}{\rm degrees}\right].
\end{equation}
Taken from Table~\ref{sources}, $D$ is the distance to the source, $\Delta_{\rm red-blue}$
the red- and blue-shifted maser separation, and $\Delta V_{\rm
  H_2O}$ the velocity separation. Furthermore, $i$ is the inclination from
the plane of the sky.  Although the distance to the IRAS~18455+0448 is
uncertain, \citet{Lewis01} argue that the luminosity of the source
suggests it is located close to the far kinematic distance of
$11.2$~kpc \cite[recalculated using the results from][]{Reid09} instead
of the near kinematic distance ($2.1$~kpc), as in that case the
luminosity would be very low for a post-AGB star. Adopting
$D=11.2~(2.1)$~kpc and an inclination of $i=45^\circ$, we find $t_{\rm
  kin}\approx67~(13)$~yr, making IRAS~18455+0448 one of the youngest
water fountain nebulae \citep{Imai07}.

The rebirth of IRAS~18455+0448 as a water fountain source, provides us
with important historical information on a water fountain
progenitor. Its small OH maser expansion velocity of $\sim6$~\kms,
implies a low main sequence mass of $\sim1.0\pm0.5$~M$_\odot$ \citep{Baud83},
which is noticeably lower than the masses derived for CO detected
water fountain sources \citep[e.g.,][]{He08, Rizzo13} and from source kinematics \citep{Imai12}.

In the context of the model proposed by \citet{Lewis01}, the OH maser
observations would indicate a direct connection between the events
leading to a gradual decrease in mass loss, which started
$\sim1000$~yr ago, and the launch of the jet $\sim70$~yr ago. Such a
relation is not obvious if the jet originates from interaction with a
binary companion during a common-envelope (CE) phase. An outflow
launched as a result from CE evolution, either due to direct envelope
ejection, an explosive dynamo-driven jet or the formation of a
circumstellar disk by shredding of the companion would act at
timescales of only a few years \citep[e.g.,][]{Nordhaus07}. The CE
outflow would also not produce an obvious relation between the smooth
exponential decay of 1612~MHz OH maser emission and the delayed
decrease of 1665/67~MHz OH maser flux.

Instead of a gradual mass loss decrease, the main cause of
the maser decay can also be, as mentioned above, the loss of pumping
efficiency instead of a decrease of OH column density. In that case, the
launch of the \water maser jet could have occurred near simultaneously to
a more rapid mass loss decrease or other event in which the dust that
produces the infrared maser pump radiation is
affected. In\citet{Lewis01}  it is however noted that there is no indication of
any infrared variability of IRAS~18455+0448.

\begin{figure*}
\centering
\resizebox{0.9\hsize}{!}{\includegraphics{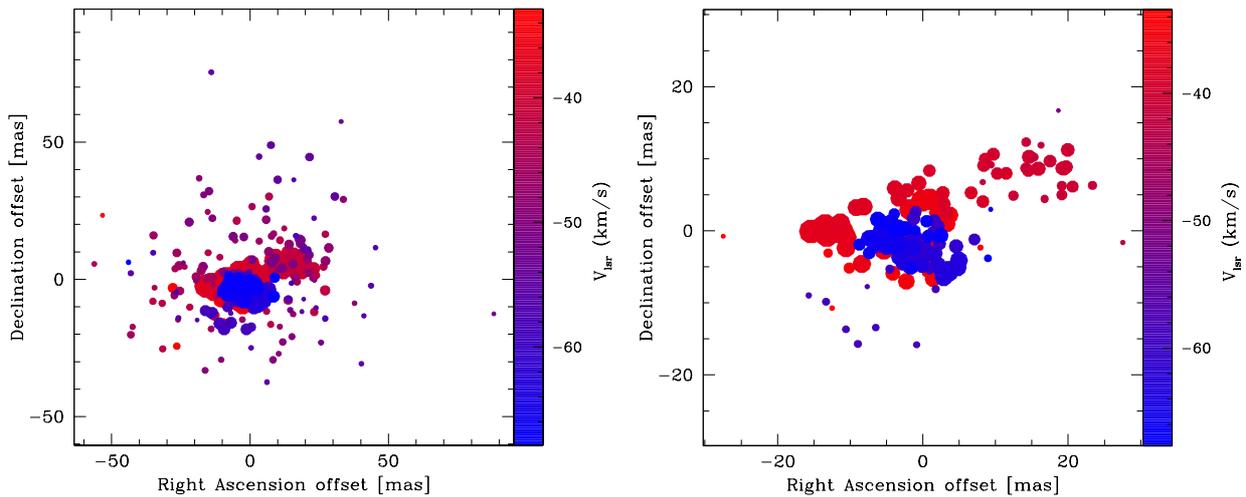}}
\caption{(left) Same as Fig.~\ref{iras18455} for IRAS~19422+3506. All
 channels with a SNR $>15$ are plotted. Positional uncertainties
 for the individuals spots range from $\sim2$~mas to $\sim30$~mas for the
 strongest and weakest plotted masers, respectively. No average
 offset is detected between the red- and blue-shifted masers. (right) Similar
 to the left but only spots with SNR $>50$ are plotted to highlight
 the northwest extension of the red-shifted masers. Individual errors 
 range from $\sim2$~mas to $\sim10$~mas.} \label{iras19422}
\end{figure*}

\subsection{Nature of the Remaining Sources}
 
None of the other sources in our sample have a \water maser velocity
range larger than that of the OH~masers. Should the \water masers
occur in a fast bipolar outflow, we would thus expect these to lie
predominantly in the plane of the sky. However, no significant offset
between the flux averaged red- and blue-shifted emission peaks is
seen. It is therefore more likely that the observed double peaked
\water maser spectra for these sources arise because of radial maser
amplification. This indicates that the sources are likely high mass-loss OH/IR stars
with a \water maser shell beyond the acceleration region, where the
largest velocity coherent maser path length is, as for the OH masers,
in the radial direction \citep[e.g.,][]{Engels96}. The multiple peaked
structure, most notable in IRAS~19422+3506, could indicate that
both radial and tangential amplification is operating in an extended maser
envelope that encompasses both accelerating material closer to the
star and constant velocity gas farther out. Alternatively, as argued
by \citet{Engels97} for the OH/IR star OH~$39.7+1.5$, the masers occur
in the same region with mode switching between tangential and radial
beaming occurring in phase with the pulsation cycle and an
accompanying changes in physical conditions. As shown in
Fig.~\ref{iras19422}, IRAS~19422+3506 is the only source besides
IRAS~18455+0448 that shows some structure within the \water maser
envelope at scales probed by our JVLA observations. While the overall
envelope appears spherical, and there is no appreciable offset between
the red- and blue-shifted emission peaks, the red-shifted emission
around $-40$~\kms~is offset by $\sim20$~mas ($\sim200$~AU) from the
rest of the red-shifted masers. Monitoring the
motion of such a clearly distinct maser cloud in the circumstellar AGB
envelope could show if the maser cloud results from a dense clumpy
mass ejection or from turbulent motions in the envelope.

\subsection{Maser Variability}

The \water masers of all sources show significant variability between
different epochs. The multi-epoch Effelsberg observations covered a
timescale of $\sim500$ days. We found that the \water masers exhibit
significant variability in flux density and spectral
characteristics. We observed the rise of new maser features that
differ by as much as $35$~\kms~in velocity from features detected in
the earlier observations \citep{Engels96} for IRAS 19422+3506. For
IRAS 18455+0448 the spectra show emission blue-shifted from the
stellar velocity at $V_{\rm LSR}\approx 34.1$~\kms~ between $V_{\rm
  LSR}=20$ and $30$~\kms~ (Fig.\ref{irasspec}, left). By our second
epoch in March 2011 the spectrum had changed within the same blue
shifted range and a weak red-shifted feature at $\sim46$~\kms~ had
appeared. One month later in April 2011, the most blue-shifted feature
halved in flux density while the red-shifted component became stronger
by a factor of a few. A further Effelsberg epoch was taken by
\citet{Yung13} in December 2011, which again showed two strong
features at $\sim19$~and $\sim46$~\kms~ and a weak third feature even
more blue-shifted at $\sim11$~\kms. Our JVLA observations from June
2013 show the now persisting red-shifted feature and show again an
even more complex blue-shifted set of maser features between
$\sim13-27$~\kms. By now, the velocity spread of the \water masers is
$>38$~\kms~ compared to the $12$~\kms~ spread of the OH masers.

In \citet{AmiriThesis}, it was found that the observed amount of
strong variability is consistent with a variability timescale of
several thousand days. This is similar to the typical OH/IR pulsation
period. For Mira variables, a similar relation between variability
timescale of the \water masers and the pulsation period was found by
\citet{Shintani08}. However, since the individual sources studied in
this work are mainly OH/IR stars with periods up to $\sim2000$ days,
we do not have optical light curves available to examine whether any
correlation exists between the stellar pulsation and \water maser
emission.

\section{Conclusions} 
\label{conclusions}

We have confirmed the previously `dead' OH/IR star IRAS~18455+0448 to
be the newest member of the class of water fountain sources. The kinematic age of IRAS~18455+0448 is $\sim70$~yr, with the main uncertainties being the yet
unknown inclination of the bipolar outflow and its distance. This is
the first time that a water fountain can be
directly linked to a previously studied OH/IR star. Based on the
originally low outflow velocity measured from the OH masers that have
now disappeared, IRAS~18455+0448 is estimated to be a fairly
low-mass star, in contrast to what is generally postulated for the
typical water fountain progenitor. The applicability of the expansion
velocity - main sequence mass relation, which yields this low mass, to
individual sources is however highly uncertain. For example
observations of CO and its isotopologues may provide further
mass estimates \citep[e.g.,][]{He08}. Furthermore, Very Long Baseline Interferometry
(VLBI) observations will be needed to derive a more accurate
distance, and constrain internal kinematics. If the original model 
describing the approaching death of IRAS~18455+0448 \citep{Lewis01} is correct, 
the launch of the bipolar outflow in recent years is linked to a longer term slow
decline of the mass loss of the water fountain progenitor. This would
be difficult to reconcile with several of the binary outflow
generation models that often act on relatively short
timescales. However, models where the exponential decrease of the OH
masers of IRAS~18455+0448 is the result of a more recent disturbance
of the mass loss or maser pumping mechanism that is at the same time
responsible for the launch of the outflow still need to be
investigated in more detail.

The remaining six sources observed with the JVLA do not show any
offset between the red- and blue-shifted masers, which would be
expected for a bipolar outflow not developed close to the line of
  sight. The double peaked nature of their \water maser spectra is
more likely the result of radial maser beaming.

 
\begin{acknowledgements} 
  WV acknowledges support by the Swedish Research Council (VR), Marie Curie Career Integration Grant
  321691 and ERC consolidator grant 614264. The research of
  NA was supported by the ESTRELA fellowship, the EU Framework 6 Marie
  Curie Early Stage Training program under contract number
  MEST-CT-2005-19669.
\end{acknowledgements} 
 


\begin{thebibliography}{99}
\bibitem[Amiri(2011)]{AmiriThesis} Amiri, N.\ 2011, Ph.D.~Thesis,
  Leiden University, https://openaccess.leidenuniv.nl/handle/1887/17981
\bibitem[Balick \& Frank(2002)]{Balick02} Balick, B., \& Frank, A.\ 2002,
\araa, 40, 439
\bibitem[Baud 
\& Habing(1983)]{Baud83} Baud, B., \& Habing, H.~J.\ 1983, \aap, 127, 73
\bibitem[Chengalur et al.(1993)]{Chengalur93} Chengalur, J.~N., 
Lewis, B.~M., Eder, J., \& Terzian, Y.\ 1993, \apjs, 89, 189
\bibitem[Eder et al.(1988)]{Eder88} Eder, J., Lewis, B.~M., 
\& Terzian, Y.\ 1988, \apjs, 66, 183 
\bibitem[Engels et al.(1986)]{Engels86} Engels, D., Schmid-Burgk, J.,
  \& Walmsley, C.~M.\ 1986, \aap, 167, 129
\bibitem[Engels \& Lewis(1996)]{Engels96} Engels, D., \& Lewis, B.~M.\
  1996, \aaps, 116, 117
\bibitem[Engels et 
al.(1997)]{Engels97} Engels, D., Winnberg, A., Walmsley, C.~M., \&
Brand, J.\ 1997, \aap, 322, 291
\bibitem[Garcia-Segura et al.(2014)]{Garcia14} Garcia-Segura, 
G., Villaver, E., Langer, N., Yoon, S.-C., \& Manchado, A.\ 2014, \apj 783, 74
\bibitem[Gray et al.(2005)]{Gray05} Gray, M.~D., Howe, D.~A., 
\& Lewis, B.~M.\ 2005, \mnras, 364, 783
\bibitem[He et al.(2008)]{He08} He, J.~H., Imai, H., Hasegawa, T.~I., Campbell, S.~W., \& Nakashima, J.\ 2008, \aap, 488, L21
\bibitem[Imai(2007)]{Imai07} Imai, H.\ 2007, IAU Symposium, 242, 279
\bibitem[Imai et al.(2012)]{Imai12} Imai, H., Chong, S.~N., He, J.-H., et al.\ 2012, \pasj, 64, 98
\bibitem[Kim et al.(2010)]{Kim10} Kim, J., Cho, S.-H., Oh, 
C.~S., \& Byun, D.-Y.\ 2010, \apjs, 188, 209
\bibitem[Kim et al.(2013)]{Kim13} Kim, J., Cho, S.-H., 
\& Kim, S.~J.\ 2013, \aj, 145, 22
\bibitem[Lewis(1997)]{Lewis97} Lewis, B.~M.\ 1997, \apjs, 109, 
489
\bibitem[Lewis et al.(2001)]{Lewis01} Lewis, B.~M., 
Oppenheimer, B.~D., \& Daubar, I.~J.\ 2001, \apjl, 548, L77
\bibitem[Likkel et al.(1992)]{Likkel92} Likkel, L., Morris, M., \&
  Maddalena, R.~J.\ 1992, \aap, 256, 581 
\bibitem[Marti-Vidal et al.(2014)]{MV14} Marti-Vidal, I., 
Vlemmings, W.~H.~T., Muller, S., \& Casey, S.\ 2014, \aap
563, 136 
\bibitem[Nakashima 
\& Deguchi(2003)]{Nakashima03} Nakashima, J.-I., \& Deguchi, S.\ 2003, \pasj, 55, 229
\bibitem[Nordhaus et al.(2007)]{Nordhaus07} Nordhaus, J., 
Blackman, E.~G., \& Frank, A.\ 2007, \mnras, 376, 599
\bibitem[P{\'e}rez-S{\'a}nchez et al.(2013)]{PerezSanchez13} 
P{\'e}rez-S{\'a}nchez, A.~F., Vlemmings, W.~H.~T., Tafoya, D., 
\& Chapman, J.~M.\ 2013, \mnras, 436, L79
\bibitem[Reid et al.(2009)]{Reid09} Reid, M.~J., Menten, 
K.~M., Zheng, X.~W., et al.\ 2009, \apj, 700, 137
\bibitem[Rizzo et al.(2013)]{Rizzo13} Rizzo, J.~R., G{\'o}mez, J.~F.,
  Miranda, L.~F., et al.\ 2013, \aap, 560, A82
\bibitem[Sahai \& Trauger(1998)]{Sahai98} Sahai, R., \& Trauger, J.~T.\ 1998, \aj, 116, 1357 
\bibitem[Shintani et al.(2008)]{Shintani08} Shintani, M., Imai, 
H., Ando, K., et al.\ 2008, \pasj, 60, 1077
\bibitem[Su{\'a}rez et al.(2008)]{Suarez08} Su{\'a}rez, O., 
G{\'o}mez, J.~F., \& Miranda, L.~F.\ 2008, \apj, 689, 430
\bibitem[te Lintel Hekkert et 
al.(1989)]{tLH89} te Lintel Hekkert, P., Versteege-Hensel, H.~A., Habing, H.~J., \& Wiertz, M.\ 1989, \aaps, 78, 399 
\bibitem[Vlemmings et al.(2006)]{Vlemmings06} Vlemmings, W.~H.~T., 
Diamond, P.~J., \& Imai, H.\ 2006, \nat, 440, 58
\bibitem[Yung et al.(2011)]{Yung11} Yung, B.~H.~K., Nakashima, 
J.-i., Imai, H., et al.\ 2011, \apj, 741, 94
\bibitem[Yung et al.(2013)]{Yung13} Yung, B.~H.~K., Nakashima, 
J.-i., Imai, H., et al.\ 2013, \apj, 769, 20

\end{thebibliography}
\end{document}